\documentclass[12pt]{article}
\usepackage{amssymb}
\usepackage{amsmath}
\usepackage{graphicx}
\begin{document}
%
%

\title{Holographic cosmology and tachyon inflation}

\author{
Neven Bili\'c \thanks{bilic@irb.hr} 
   \\
{\it Division of Theoretical Physics, Rudjer Bo\v{s}kovi\'{c} Institute,}
\\
{\it Zagreb, Croatia}
}

\maketitle

\begin{abstract}
After a brief introduction to the AdS/CFT holography a tachyon  inflation 
will be discussed in the framework of holographic cosmology. 
The model is based on a holographic braneworld  scenario  
with an effective  tachyon field on a D3-brane located at the holographic bound of an asymptotic ADS$_5$ bulk.
 \end{abstract}

\section{Introduction}
Branewarld cosmology is based on the  scenario in which matter is confined 
on a brane moving in the higher dimensional bulk
with only gravity allowed to propagate in the bulk  \cite{arkani,antoniadis,randall1,randall2}.
The brane can be placed, e.g., at the boundary of
a 5-dim asymptotically anti de Sitter (AdS) space-time.
We shall refer to this type of braneworld as the holographic braneworld 
\cite{apostolopoulos,bilic1}.

It is important to  stress 
that AdS space appears naturally in the context of M/string theory.
The point is that pure supergravity theories, without
additional supersymmetric matter in arbitrary dimensions, necessarily satisfy the strong energy condition \cite{gibbons}.
After compactification of 10 or 11-dimensional theory down to 4+1-dimensional 
spacetime X, the stress tensor in X also satisfies the strong energy condition.
As a consequence, if we wish X to be a maximally symmetric spacetime,  it must be an Einstein space with
non-positive cosmological constant, i.e., X must be either Minkowski or anti de Sitter.

In the study of the holographic braneworld  a crucial property of an asymptotically AdS bulk
is that  AdS space is dual to a conformal field theory at
its boundary.  The so called anti de Sitter/conformal field theory (AdS/CFT) correspondence \cite{maldacena}
reflects an obvious symmetry relationship:
on the bulk side, AdS$_5$ is a maximally symmetric solution to Einstein’s
equations with negative cosmological constant with 
the symmetry group  AdS$_5$ $\equiv$ SO(4,2);
on the boundary side,
the boundary conformal field theory is invariant under
conformal transformations: Poincar\'{e} + dilatations + special
conformal transformation. These transformations constitute the conformal group 
which is locally isomorphic to the SO(4,2) symmetry group.

Our aim here is to study a model of tachyon inflation  
in the framework of   holographic cosmology. 
In tachyon inflation models \cite{fairbairn,frolov,sami,cline,steer}
inflation is driven by the tachyon field originating in string theory.
In particular, our model is based on 
a holographic braneworld  scenario  with an effective  tachyon field on the 
D3-brane located at the holographic boundary of an asymptotic ADS$_5$ bulk.
This scenario is partly based on our previous works \cite{bilic2,bilic3,bilic4}.

The remainder of the paper is organized as follows. We begin by section \ref{adscft} 
in which we give brief introduction to
the AdS/CFT correspondence including the holographic renormalization (section \ref{renormalization}) and a derivation
of holographic cosmology (section \ref{cosmology}). In this section we add a subsection \ref{gauss-bonnet} in which we comment on a modified Gauss-Bonnet gravity
in relation to holographic cosmology.
In section \ref{inflation} we propose a model for tachyon inflation and present preliminary results. 
Conclusions and outlook are given
in section \ref{conclusions}.

 \begin{figure}[t]
\begin{center}
\includegraphics[width=\textwidth,trim= 0 3cm 0 0]{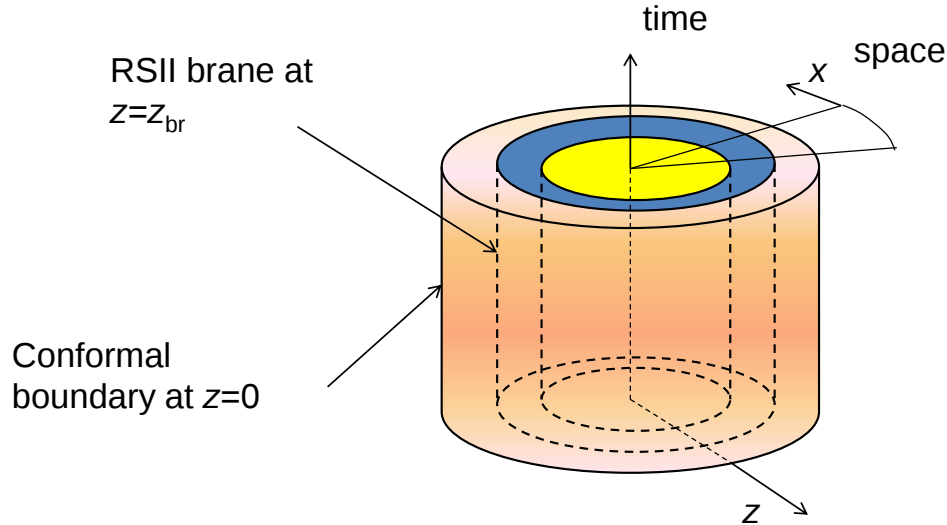}
\caption{Illustration of the AdS$_5$ bulk with two branes: Randall-Sundrum (RSII) brane located at $z=z_{\rm br}$  
and the holographic brane at $z=0$.
}
\label{fig1}
\end{center}
\end{figure}

\section{AdS/CFT and holographic cosmology}
\label{adscft}

 A general asymptotically AdS$_5$  metric in Fefferman-Graham coordinates \cite{fefferman}
is of the form
\begin{equation}
ds^2=G_{ab}dx^adx^b =\frac{\ell^2}{z^2}\left( g_{\mu\nu} dx^\mu dx^\nu -dz^2\right),
 \label{eq3001}
\end{equation}
where the length scale $\ell$ is the AdS curvature radius and
we use the Latin alphabet for 4+1  and the Greek alphabet for 3+1 spacetime indices. 
 Consider a 5-dim bulk action in AdS$_5$ background
\begin{equation}
S_{(5)}[\Phi]=\int d^5x \sqrt{G}\mathcal{L}_{(5)}(\Phi,G_{ab}).
 \label{eq100}
\end{equation}
Given an induced metric $h_{\mu\nu}$ on the boundary and a boundary value $\Phi(x,z=0)\equiv \phi(x)$,  
the bulk field $\Phi$  is completely determined by its field equations
obtained from the variational principle
\begin{equation}
\frac{\delta S_{(5)}}{\delta \Phi}=0.
 \label{eq101}
\end{equation}
A solution $\Phi[\phi,h]$ to this equation can be used to define a functional
\begin{equation}
S[\phi,h] = S_{(5)}^{\rm sh}\left[\Phi[\phi,h]\right],
\end{equation}
where
$S_{(5)}^{\rm sh}\left[\Phi[\phi,h]\right]$
is the on shell bulk action.
Then the AdS/CFT conjecture states that $S[\phi,h]$ can be identified with 
the generating
functional of a conformal field theory (CFT) on the boundary, i.e.,
\begin{equation}
S[\phi,h] \equiv \ln \int d\psi \exp \left\{ -\int d^4x\sqrt{|h|} 
\left[ \mathcal{L}^{\rm CFT}(\psi(x)) - O(\psi(x))\phi(x)      \right]\right\},
\end{equation}
where $\mathcal{L}^{\rm CFT}$ is a CFT Lagrangian, $O(\psi)$ are CFT operators of
dimension $\Delta$, and 
the boundary fields serve as sources for CFT operators.
In this way the CFT correlation functions can be calculated as functional
derivatives of the on-shell bulk action, e.g.,
\begin{equation}
\frac{\delta^2 S_{(5)}}{\delta \phi(x)\delta \phi(y)}
=\langle O(\psi(x))O(\psi(y))\rangle-\langle O(\psi(x))\rangle \langle O(\psi(y))\rangle .
 \label{eq104}
\end{equation}
Similarly, the induced metric $h_{\mu\nu}$ serves as the source for the 
 the stress tensor operator of the
dual CFT so that its vacuum expectation value is obtained as
\begin{equation}
\frac{1}{2\sqrt{|h|}}\frac{\delta S_{(5)}}{\delta h^{\mu\nu}}
=\langle T^{\rm CFT}_{\mu\nu}\rangle .
 \label{eq105}
\end{equation}

\subsection{Holographic renormalization}
\label{renormalization}

Consider next the bulk action with only gravity in the bulk
 \begin{equation} 
S_{(5)}[G]=\frac{1}{8\pi G_5} \int d^5x \sqrt{G} 
\left[-\frac{R^{(5)}[G] }{2} -\Lambda _5 \right] ,
\label{eq106} 
\end{equation}
where $G_5$ is the five-dimensional gravitational constant and $\Lambda_5$ is
the bulk cosmological constant related to the AdS curvature radius as 
$\Lambda_5=-6/\ell^2$.
The on-shell action is IR divergent 
and must be regularized and renormalized.
The asymptotically AdS metric near $z=0$ can be expanded as
\begin{equation}
  g_{\mu\nu}(z,x)=g^{(0)}_{\mu\nu}(x)+z^2 g^{(2)}_{\mu\nu}(x)+z^4 g^{(4)}_{\mu\nu}(x) +\cdots .
  \label{eq107}
 \end{equation}
 Explicit expressions for $g^{(2n)}_{\mu\nu}$
in terms of arbitrary $g^{(0)}_{\mu\nu}$
can be found in Ref.\ \cite{deharo}.
We regularize the action by placing a brane (RSII brane)
near the
boundary, i.e., at $z=\epsilon \ell$, $\epsilon\ll 1$, so that the induced metric on the brane is
 \begin{equation}
  h_{\mu\nu}=\frac{1}{\epsilon^2}g_{\mu\nu}(\epsilon \ell,x)
=\frac{1}{\epsilon^2}\left(g^{(0)}_{\mu\nu}+\epsilon^2\ell^2 g^{(2)}_{\mu\nu} 
+\epsilon^4\ell^4 g^{(4)}_{\mu\nu}
+ \cdots \right) .
 \end{equation}
  The bulk splits in two regions: $0\leq z < \epsilon \ell$ and $\epsilon \ell\leq z < \infty $. 
  We can either
discard the $0\leq z  < \epsilon\ell$ region (one-sided regularization) or invoke the Z$_2$
symmetry and identify two regions (two-sided regularization). For
simplicity we shall use the one-sided regularization. The regularized on shell
bulk action is \cite{deharo2}
  \begin{equation} 
S_{(5)}^{\rm reg}[h] =\frac{1}{8\pi G_5} \int\limits_{z\geq \epsilon\ell} d^5x \sqrt{G} 
\left[-\frac{R^{(5)} }{2} -\Lambda _5 \right] 
+S_{\rm GH}[h]+S_{\rm br}[h],
\label{eq001} 
\end{equation}
 where $S_{\rm GH}$ is the Gibbons-Hawking boundary  term  and the brane action is given by
 \begin{equation} 
S_{\rm br}[h] =\int d^{4}x\sqrt{|h|} (-\sigma + \mathcal{L}_{\rm matt}[h]),
\label{eq1005}
\end{equation} 
where $\sigma$ is the brane tension and the Lagrangian $\mathcal{L}_{\rm matt}$ 
describes matter on the brane.
The renormalized  action is obtained by adding
counter-terms and taking the limit $\epsilon\rightarrow 0$
\begin{equation} 
S^{\rm ren}[h]=S_{(5)}^{\rm reg}[h]+S_1[h]+S_2[h]+S_3[h],
\label{eq1007} 
\end{equation}
 The necessary counter-terms are \cite{deharo,hawking}
\begin{equation} 
S_1[h]=-\frac{6}{16\pi G_5\ell}\int d^4x \sqrt{|h|} , 
\label{eq4001} 
\end{equation}
\begin{equation} 
S_2[h]=-\frac{\ell}{16\pi G_5}\int d^4x \sqrt{|h|}\left(-\frac{R[h]}{2} \right) ,
\label{eq4002} 
\end{equation}
\begin{eqnarray} 
S_3[h]=-\frac{\ell^3}{16\pi G_5}\int d^4x \sqrt{|h|}\frac{\log\epsilon}{4}
 {\bigg(} R^{\mu\nu}[h]R_{\mu\nu}[h] 
\left. 
 -\frac13 R^2[h] \right).
\label{eq4003} 
\end{eqnarray}
The last term is scheme dependent and its integrand is proportional to the 
holographic conformal anomaly
\cite{henningson}.
Now we demand that the variation with respect to the induced metric $h_{\mu\nu}$ of
the regularized on shell bulk action (RSII action)  vanishes, i.e., we require
\begin{equation}
 \delta S_{(5)}^{\rm reg}[h]=0,
\end{equation}
which may be expressed as
\begin{eqnarray}
\!\!&&
\delta\left[S^{\rm ren}-  S_3-\left(\sigma-\frac{3}{8\pi G_5\ell}\right)\int\! d^4x \sqrt{|h|}
+\int\! d^4x \sqrt{|h|}\mathcal{L}_{\rm matt}
\right. 
\nonumber\\
\!\!\!\!\!&&\left.
-\frac{\ell}{16\pi G_5}\int\! d^4x \sqrt{|h|}\frac{R[h]}{2} \right]\!=\!0. 
\label{eq4005} 
\end{eqnarray}
The third term gives the contribution to the cosmological constant and may be eliminated 
by imposing the RSII fine-tuning condition 
\begin{equation} 
\sigma= \frac{3}{8\pi G_5\ell}.
\label{eq0012} 
\end{equation}
The variation  of the scheme dependent $S_3$ may be combined with the first term so that
\begin{equation}
  \frac{2}{\sqrt{-h}}
 \frac{\delta S^{\rm ren}}{\delta h^{\mu\nu} }
-\frac{2}{\sqrt{-h}}
 \frac{\delta S_3}{\delta h^{\mu\nu}}=\langle T^{\rm CFT}_{\mu\nu}\rangle,
\label{eq3004}
\end{equation}
according to the AdS/CFT prescription (\ref{eq105}).
The net effect of $\delta S_3$ is that it cancels the $\Box R$ term
in the conformal anomaly \cite{kiritsis} so  the trace of the CFT stress tensor simply reads
\begin{equation}
\langle {T^{\rm CFT}}^\mu_\mu\rangle =-\frac{\ell^3}{64\pi G_5}
 \left( R^{\mu\nu}R_{\mu\nu} -\frac13 R^2 \right).  
\label{eq3122}
\end{equation}

The variation equation (\ref{eq4005})
 yields  four-dimensional Einstein's equations on the boundary 
\begin{equation}
R_{\mu\nu}- \frac12 R g^{(0)}_{\mu\nu}= 8\pi G_{\rm N} (\langle T^{\rm CFT}_{\mu\nu}\rangle +T^{\rm matt}_{\mu\nu}),
 \label{eq3002}
\end{equation}
where $R_{\mu\nu}$ is the Ricci tensor associated with the metric $g^{(0)}_{\mu\nu}$
and the energy energy-momentum tensor 
\begin{equation}
{T^{\rm matt}}^{\mu}_{\nu}=\mbox{diag}(\rho, -p,-p,-p) 
\label{eq3010}
\end{equation}
corresponds to the Lagrangian $\mathcal{L}_{\rm matt}$ and describes matter on the brane.
Thanks to the AdS/CFT correspondence, 
the vacuum expectation value $\langle T^{\rm CFT}_{\mu\nu}\rangle$  is obtained in terms  
of quantities related to the bulk metric \cite{deharo}
\begin{eqnarray}
  \langle T^{\rm CFT}_{\mu\nu}\rangle=
 \frac{\ell^3}{4\pi G_5}\left\{
\frac18 \left[({\rm Tr} g^{(2)})^2-{\rm Tr} (g^{(2)})^2\right]g^{(0)}_{\mu\nu}
 +\frac12 (g^{(2)})^2_{\mu\nu}-\frac14 {\rm Tr} g^{(2)}g^{(2)}_{\mu\nu}
 -g^{(4)}_{\mu\nu} \right\}.
 \label{eq3106}
\end{eqnarray}
In the next section we will specify the boundary metric in the FRW form and
derive the evolution equations on the holographic brane. 

\subsection{Holographic cosmology}
\label{cosmology}
Here  we outline a derivation of the Friedmann equations on the holographic brane following
Refs.\  \cite{apostolopoulos,bilic1}.
For this purpose it is convenient 
to represent the bulk metric 
in AdS-Schwarzschild static coordinates \cite{birmingham1}
\begin{equation}
ds_{(5)}^2= 
f(r) d\tau^2- \frac{dr^2}{f(r)} -r^2 d\Omega_k^2, 
\label{eq3202}
\end{equation}
where 
\begin{equation}
f(r)=\frac{r^2}{\ell^2}+k -\mu \frac{\ell^2}{r^2}
\label{eq3225}
\end{equation}
and
\begin{equation}
d\Omega^2_k=d\chi^2+\frac{\sin^2(\sqrt{k}\chi)}{k}(d\vartheta^2+\sin^2 \vartheta d\varphi^2)
\label{eq1004}
\end{equation}
is the spatial line element for a 
closed ($k=1$), open hyperbolic ($k=-1$), or open flat ($k=0$) space.
The dimensionless parameter $\mu$ is  related to the black-hole mass via \cite{myers,witten}
\begin{equation}
\mu=\frac{8G_5 M_{\rm bh}}{3\pi \ell^2}.
 \label{eq3105}
\end{equation}
Starting from (\ref{eq3202}) we make a coordinate transformation 
\begin{equation}
 \tau=\tau(t,z), \quad  r=r(t,z),
 \label{eq204}
\end{equation}
such that the line element in new coordinates  takes the form
\begin{equation}
ds_{(5)}^2=\frac{\ell^2}{z^2}\left( 
{\mathcal{N}}^2(t,z)dt^2- {\mathcal{A}}^2(t,z) d\Omega_k^2-dz^2 
\right) ,
 \label{eq102}
\end{equation}
where 
\begin{equation}
 {\mathcal{A}}^2(t,z)=\frac{z^2}{\ell^2}r^2(t,z)   
 \label{eq103}
\end{equation}
and $\mathcal{N}$ is fixed by the requirement that the off diagonal elements 
of the transformed metric vanish.
Next, imposing the boundary conditions at z=0:
\begin{equation}
\mathcal{N}(t,0)=1, \quad {\mathcal{A}}(t,0) = a(t), 
 \label{eq302}
\end{equation}
we obtain the induced metric at the boundary in the FRW
form
\begin{equation}
ds^2=g^{(0)}_{\mu\nu}dx^\mu dx^\nu =dt^2 -a^2(t) d\Omega_k^2 .
 \label{eq3201}
\end{equation}
Solving Einstein’s equations in the bulk one finds \cite{apostolopoulos}
\begin{equation}
\mathcal{A}^2=a^2\left[
1-\left(H^2+\frac{k}{a^2}\right) \frac{z^2}{4}\right]^2
+ \frac14 \frac{\mu z^4}{a^2},
 \label{eq3103}
\end{equation}
\begin{equation}
{\mathcal{N}}=\frac{\dot{\mathcal{A}}}{\dot{a}},
 \label{eq3104}
\end{equation}
where $H=\dot{a}/a$ is the Hubble expansion rate  on  the $z=0$ boundary.
Comparing the exact solution with the expansion (\ref{eq107})
we can extract $g^{(2)}_{\mu\nu}$ and $g^{(4)}_{\mu\nu}$.
Then, using  (\ref{eq3106}) we find the expression 
for $T^{\rm CFT}$ in the special case of the boundary metric (\ref{eq3201})
\begin{equation}
 \langle T^{\rm CFT}_{\mu\nu}\rangle = t_{\mu\nu}+
\frac14 \langle {T^{\rm CFT}}^\alpha_\alpha\rangle g^{(0)}_{\mu\nu} .
 \label{eq3107}
\end{equation}
The second term on the right-hand side  corresponds to the conformal anomaly 
\begin{equation}
 \langle {T^{\rm CFT}}^\alpha_\alpha\rangle = 
\frac{3\ell^3}{16\pi G_5}\frac{\ddot{a}}{a} \left(H^2+\frac{k}{a^2}\right) .
 \label{eq3027}
\end{equation}
The first term on the right-hand side of (\ref{eq3107}) is
 a traceless tensor the nonvanishing components of which are
\begin{equation}
 t_{00}=-3 t^i_i =\frac{3\ell^3}{64\pi G_5 }
\left[ \left(H^2+\frac{k}{a^2}\right)^2+\frac{4\mu}{a^4} 
-\frac{\ddot{a}}{\dot{a}}\left(H^2+\frac{k}{a^2}\right)\right] .
 \label{eq3108}
\end{equation}
Hence, apart from the conformal anomaly, the CFT dual to the time
dependent asymptotically AdS$_5$ metric (\ref{eq102}) is a conformal fluid with the
equation of state $p_{\rm CFT}=\rho_{\rm CFT}/3$,
where $\rho_{\rm CFT}=t_{00}$, $p_{\rm CFT}=-t^i_i$.

Next, using the effective Einstein equations (\ref{eq3002})  we obtain the holographic
Friedmann equation \cite{apostolopoulos,kiritsis} which for a spatially flat geometry ($k=0$)
takes the form
\begin{equation}
 H^2-\frac{\ell^2}{4}H^4=
 \frac{8\pi G_{\rm N}}{3}\rho+\frac{4\mu}{a^4}. 
 \label{eq3110}
\end{equation}
Hence, one finds a deviation from the standard cosmology
in two terms: the term proportional to $H^4$ on the left-hand side and the 
last term on the right-hand side, the so called ``dark radiation''.
Combining (\ref{eq3110}) and the energy conservation equation
\begin{equation}
 \dot{\rho}+3H(p+\rho)=0,
\end{equation}
one can easily derive the second Friedmann equation on the holographic brane
\begin{equation}
 \dot{H} \left(1-\frac{\ell^2}{2}H^2\right)=
 -4\pi G_{\rm N}(p+\rho). 
 \label{eq3111}
\end{equation}

It is worth mentioning that holographic type cosmologies have been studied in other contexts
\cite{barvinsky,lidsey,viaggiu,gao,kiritsis2,binetruy}.
In particular, a modified Gauss-Bonnett gravity \cite{cognola}
in a cosmological context
 leads in a special case to the equation of the form (\ref{eq3110}) with $\mu=0$.
It is instructive to briefly elaborate on this special form of modified Gauss-Bonnet gravity
following C.~Gao \cite{gao}.

\subsection{Connection with modified Gauss-Bonnet gravity}
\label{gauss-bonnet}
A modified Gauss-Bonnet gravity belongs to a class of modified gravity models in which
the gravitational action is a general function of two variables: the Ricci scalar $R$ and  the Gauss-Bonnet invariant 
\begin{equation}
\mathcal{G}=R^2 -4R^{\mu\nu}R_{\mu\nu}+R^{\mu\nu\rho\sigma}R_{\mu\nu\rho\sigma}.
 \label{eq1001}
\end{equation}
The total action in such a model can be written in the form
\begin{equation}
S=\int d^4x\sqrt{-g}\left[
\frac{1}{16\pi G_{\rm N}}\left(-R+f(R,\mathcal{G})\right)+\mathcal{L}_{\rm matt}
\right],
 \label{eq1002}
\end{equation}
where $f$ is a smooth function.
This class of models was shown to be ghost free \cite{comelli,navarro}.
In a cosmological context  it is natural to 
require in addition that the second Friedmann equation is linear
in $\dot{H}$. Then it follows \cite{gao}  that  $f$ must be a function of only one variable $f = f(J)$ 
 where
\begin{equation}
J= \frac{1}{\sqrt{12}}
\left(-R+\sqrt{R^2-6\mathcal{G}}
\right)^{1/2}.
 \label{eq1003}
\end{equation}
For a spatially flat metric one finds 
$J=\dot{a}/a\equiv H$
and the first Friedmann equation
takes the form
\begin{equation}
 H^2+\frac16 \left( f(H) - H\frac{df}{dH}\right)
 =\frac{8\pi G_{\rm N}}{3}\rho .
 \label{eq1006}
\end{equation}
Hence, the left hand side is a function of $H$ only and the second Friedmann equation
will be linear in $\dot{H}$. In particular, for
\begin{equation}
f=\frac12 \ell^2 H^4 ,
 \label{eq1008}
\end{equation}
the Friedman equation takes the holographic form (\ref{eq3110}) with $\mu=0$.
Thus, the holographic cosmology is reproduced in a modified Gauss-Bonnet gravity with action
\begin{equation}
S=\int d^4x\sqrt{-g}\left[
\frac{1}{16\pi G_{\rm N}}\left(-R+\frac{\ell^2}{288}\left(\sqrt{R^2-6\mathcal{G}}-R\right)^2\right)
+\mathcal{L}_{\rm matt}
\right].
 \label{eq1009}
\end{equation}

\section{Tachyon inflation on the holographic brane}
\label{inflation}
Here we give a brief recapitulation of the results which will be presented in more detail 
elsewhere  \cite{bilic}.
We start from a general tachyon Lagrangian
\begin{equation}
{\mathcal{L}} = -\ell^{-4} V(\theta/\ell)
\sqrt{1-g^{\mu\nu}\theta_{,\mu}\theta_{,\nu}}  \, ,
 \label{eq000}
\end{equation}
where the field $\theta$ is of dimension of length and the AdS$_5$ curvature radius $\ell$  
fixes the scale. The potential $V$ is an arbitrary non-increasing function.
Assuming a spatially flat FRW metric on the holographic brane we solve the field equations 
augmented by the holographic Friedman equation (\ref{eq3110}) with $\mu=0$, 
\begin{equation}
H^2-\frac{\ell^2}{4}H^4=\frac{8\pi G_{\rm N}}{3} \rho ,
 \label{eq115}
\end{equation}
where 
$\rho=\ell^{-4}V /\sqrt{1-\dot{\theta}^2}$.
Solving (\ref{eq115}) as a quadratic equation for $H^2$ we find
\begin{equation}
H^2=\frac{2}{\ell^2}\left(1\pm\sqrt{1-\frac{8\pi G_{\rm N}}{3}\ell^2\rho}\right).
 \label{eq125}
\end{equation}
Demanding that this equation reduces to the standard Friedmann
equation in the low density limit, i.e., in the limit when
$
G_{\rm N}\ell^2\rho \ll 1
$,
we can discard the ($+$) sign solution. Then  
from (\ref{eq125}) it follows that
$H^2$ can vary between zero and its maximal value $H_{\max}^2=2/\ell^2$ corresponding to the maximal energy density
$\rho_{\rm max}= 3/(8\pi G\ell^2)$ \cite{bilic1,delcampo}.
If in addition we assume  no violation of the weak energy condition $\rho\geq 0$,
$p+\rho\geq 0$, 
 the expansion rate will, according to (\ref{eq3111}), be a monotonously decreasing function of time.
The universe evolution starts from $t=0$ with an initial $H_{\rm i} \leq H_{\rm max}$ 
 with energy density and cosmological scale both finite.
Hence, as already noted in Ref. \cite{gao}, in the  modified cosmology described
by the Friedmann equation (\ref{eq115}) 
the Big Bang singularity is avoided!

In the following we will examine a simple exponential potential $V=e^{-\omega x}$ 
where $\omega$ is a free dimensionless parameter.
This potential has been extensively exploited in the tachyon literature
\cite{sami,cline,steer,sen2}.

\begin{figure}[ht]
\begin{center}
\includegraphics[width=0.7\textwidth,trim= 0 0 0 0]{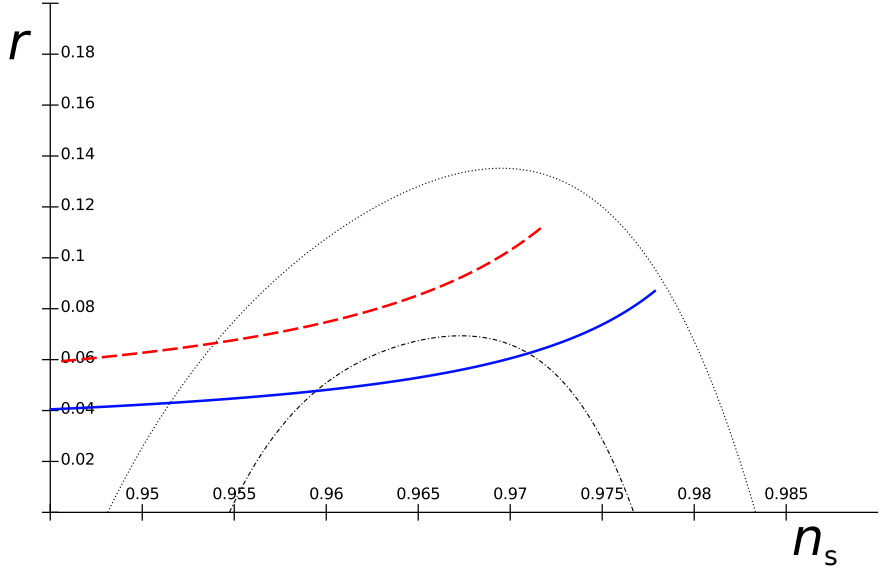}
\caption{$r$ versus $n_{\rm s}$  for fixed $N=70$ (dashed red line) and
$N=90$ (full blue line) and varying initial values $H_i^2$ ranging from
0 to 2$\ell^{-2}$ along the lines. The parameter $\omega$ is also varying  
in view of  (\ref{eq111}). The dash-dotted and dotted lines denote the Planck contours of 
the one and two $\sigma$ constraints, respectively.
} 
\label{fig4}
\end{center}
\end{figure}

Tachyon inflation is based upon the slow evolution of $\theta$ with the slow-roll
conditions 
\begin{equation}
 \dot{\theta}^2\ll 1, \quad |\ddot{\theta}| \ll 3H \dot{\theta}.
 \label{eq4101}
\end{equation}
Then, during inflation we find
\begin{equation}
H^2\ell^2\simeq 2(1-\sqrt{1-\kappa^2V/3}),
 \label{eq121}
\end{equation}
where $\kappa^2=8\pi G/\ell^2$.
Using this we can calculate the so called number of e-folds defined as
\begin{equation}
N\equiv \int_{t_{\rm i}}^{t_{\rm f}} Hdt, 
\end{equation}
where the subscripts i and f denote respectively the beginning and the end of inflation.
 This can be calculated explicitly 
yielding an expression that relates the initial expansion rate  $H_{\rm i}$ to  $\omega$ and  $N$ 
\begin{equation}
N=\frac{12}{\omega^2}\left[
\sqrt{1-\frac{\omega^2}{3}}
-1+\frac{H_{\rm i}^2\ell^2}{2}+\ln \left(2-\frac{H_{\rm i}^2\ell^2}{2}\right)
-\ln\left(1+\sqrt{1-\frac{\omega^2}{3}}   \right)
\right].
 \label{eq111}
\end{equation}

The slow-roll inflation parameters   
\begin{equation}
\varepsilon_1\equiv-\frac{\dot{H}}{H^2}=\frac{\omega^2 (4-H^2\ell^2)}{6H^2\ell^2(2-H^2\ell^2)}, \quad
\varepsilon_2\equiv\frac{\dot{\varepsilon_1}}{H\varepsilon_1}=
2\varepsilon_1\left(1-\frac{2H^2\ell^2}{(2-H^2\ell^2)(4-H^2\ell^2)}\right),
 \label{115}
\end{equation}
are related to  
the tensor-to-scalar ratio $r$ and the scalar spectral index $n_{\rm s}$
defined by
\begin{equation}
 r=\frac{\mathcal{P}_{\rm T}}{\mathcal{P}_{{\rm S}}}, \quad
 n_{\rm s}= \frac{d\ln \mathcal{P}_{{\rm S}}}{d\ln q},
 \label{eq3006}
 \end{equation}
where $\mathcal{P}_{{\rm S}}$ and $\mathcal{P}_{\rm T}$ are the power spectra of scalar
and tensor perturbations, respectively,
 evaluated at the horizon, i.e., for  a wave-number satisfying $q=aH$. 
 Combining the previously developed techniques for calculating the power spectra
 \cite{frolov,steer,delcampo,hwang} applied to the tachyon fluid we find
 at the lowest order in $\varepsilon_1$ and $\varepsilon_2$
\begin{equation}
r=16\varepsilon_1 \left(1-\frac{H^2\ell^2}{2}\right)  \left(1-C\varepsilon_2-\frac{4-2H^2\ell^2}{12-3H^2\ell^2}\varepsilon_1\right),
\label{eq128}
\end{equation}
\begin{equation}
n_{\rm s}=1+\left(-3+\frac{2-3H^2\ell^2}{2-H^2\ell^2}\right)\varepsilon_1-\varepsilon_2 ,
 \label{eq129}
\end{equation}
where $C=-2+\ln 2 +\gamma\simeq -0.72$ and $\gamma$ is the Euler constant.
These expressions  substantially deviate from
the lowest order results in the canonical scalar inflation   
and  standard tachyon inflation \cite{steer}.
A comparison of our preliminary results  
 with  Planck 2015 \cite{planck2015} data (Fig.\ \ref{fig4})  
 as well as with the most recent Planck data\footnote{The Planck 2018 data were released after the delivery of this talk.} 
 \cite{planck2018} shows a reasonable agreement.
 
\section{Conclusions and outlook}
\label{conclusions}
In this brief review we have  discussed a holographic cosmology in a braneworld scenario
applied to tachyon inflation.
We have shown that the slow-roll equations of the tachyon
inflation with exponentially attenuating potential on the
holographic brane are quite distinct from those of the standard
tachyon inflation with the same potential.
The presented results obtained in the slow roll approximation
are preliminary. What remains to be done is to solve the exact
equations numerically for the same potential and for various
other potentials that have been exploited in the literature.

 \section*{Acknowledgments}

This work  has been  supported by the H2020 CSA Twinning project No.\ 692194, ``RBI-T-WINNING''
and partially supported by the ICTP - SEENET-MTP project NT-03 Cosmology - Classical and Quantum Challenges
and by the STSM CANTATA-COST grant. 
The author acknowledges hospitality of the Physics Department, University of Nis, Serbia.
%


\end{document}